# Intrinsic Electronic Properties of BN Encapsulated, van der Waals Contacted MoSe$_2$ FETs

Yinjiang Shao (邵印江)[1], Jian Zhou (周健)[1], Ning Xu (徐宁)[1], Jian Chen (陈健)[1], Kenji Watanabe (渡邊賢司)[2], Takashi Taniguchi (谷口尚)[2], Yi Shi (施毅)[1], Songlin Li (黎松林)[1,*]

[1] School of Electronic Science and Engineering, National Laboratory of Solid-State Microstructures, and Collaborative Innovation Center of Advanced Microstructures, Nanjing University, Nanjing 210023, China
[2] National Institute for Materials Science, Tsukuba, Ibaraki 305-0044, Japan

*Email: sli@nju.edu.cn

**ABSTRACT**

Two-dimensional (2D) semiconductors have attracted considerable interest for their unique physical properties. Here, we report the intrinsic cryogenic electronic transport properties in few-layer MoSe$_2$ field-effect transistors (FETs) that are simultaneously van der Waals contacted with gold electrodes and are fully encapsulated in ultraclean hexagonal boron nitride dielectrics. The FETs exhibit electronically favorable channel/dielectric interfaces with low densities of interfacial traps ($< 10^{10}$ cm$^{-2}$), which lead to outstanding device characteristics at room temperature, including a near-Boltzmann-limit subthreshold swings (65 mV/dec), a high carrier mobility (68 cm$^2$ V$^{-1}$ s$^{-1}$), and a negligible scanning hysteresis ($< 15$ mV). The dependence of various contact-related quantities on temperature and carrier density are also systematically characterized to understand the van der Waals contacts between gold and MoSe$_2$. The results provide insightful information on the device physics in van der Waals contacted and encapsulated 2D FETs.

Two-dimensional (2D) van der Waals materials have been widely investigated for their unique atomic structure that enables emerging physical phenomena and nanoelectronics. On the one hand, the isolation and on-demand reassembly of various 2D materials together have led to the observation of intriguing phenomena, including indirect-direct bandgap transition,[1] chiral energy valleys,[2] interlayer excitons,[3] and restructured Morié quantum sates.[4–7] On the other hand, the atomically thin semiconductors that feature a high carrier mobility ($\mu$) are also considered promising candidates for nanoelectronics at technological nodes beyond silicon.[8–10] In spite of these attractive prospects, however, 2D semiconductors generally fall short of electronic and environmental instability, because all of the lattice atoms are exposed to external surroundings, which renders them susceptible to interfacial absorbates, charge impurities, and dangling bonds of contacted materials and, normally, results in long-term degradation in electronic performance.[11] Thus, appropriate device fabrication techniques that integrate interface passivation and electrically Ohmic contact are highly sought after.

The most prototypical van der Waals semiconductors can be attributed to the six transition-metal dichalcogenides (TMDCs) that share a general chemical formula $MX_2$ (M= Mo and W; X= S, Se, and Te). Among them, $MoSe_2$ remains rarely explored for its relatively low carrier mobility due to its high effective mass ~0.8 $m_0$,[12] where $m_0$ denotes the effective mass of carries. $MoSe_2$ has an indirect bandgap of 1.1 eV in bulk and a direct one of 1.58 eV in monolayer.[13,14] Although it has been widely used as FET channels,[15,16] photodetectors,[17] excitonic reflectors,[18] and spin functional elements,[19] the intrinsic electronic transport behavior, in particular at cryogenic temperatures, of few-layer $MoSe_2$ remains largely unexplored and only little literature is available.[20]

In this work, we report the simultaneous integration of few-layer $MoSe_2$, via ultraclean BN encapsulation and van der Waals contacts,[21] and the intrinsic electronic properties of the well-passivated devices. The high interface quality results in low densities of interfacial traps, negligible scanning hysteresis (< 15 mV), near-Boltzmann-limit subthreshold swings (~65 mV/dec), and a relatively high carrier mobility (68 $cm^2V^{-1}s^{-1}$) at room temperature. Analyses on the van der Waals Au/$MoSe_2$ contacts reveal a low effective thermal injection barrier of 68 meV and a minimal contact resistance of 4.5 k$\Omega$·µm. Nonmonotonic dependence of contact resistance ($R_C$) on temperature ($T$) is observed, as a result of the competition of the opposite trends between specific contact resistivity ($\rho_C$) and semiconductor channel resistivity ($R_S$).

In order to achieve ultraclean interfaces and van der Waals contacts for the $MoSe_2$ channels, we adopted a dry-pickup-transfer technique to stack the channel and top dielectric layers.[22] Figure 1a shows the schematic illustration for all the layers that comprise the BN encapsulated MoSe$_2$ FET, where the MoSe$_2$ channel is van der Waals contacted with the metallic electrodes (3 nm Ni/12 nm Au) and these two layers are sandwiched by two ultraclean BN layers. To facilitate the fabrication process, the contact electrodes with designed patterns are directly defined onto the bottom BN layer by conventional electron beam lithography, followed by thermal annealing in a mixed Ar/H$_2$ atmosphere at 250 °C for 2 h to remove the possible resist residues accumulated during lithography.[21] The final device and wiring configurations are shown in Figure 1b.

For clarify, Figures 1c−1g show the optical images taken at different fabrication stages. At first, typical Hall electrodes are fabricated on the bottom BN support (Figure 1c). Then, a MoSe$_2$ channel is initially exfoliated onto a 285 nm SiO$_2$/Si substrate (Figure 1d), followed by a polymer-free pickup of the MoSe$_2$ channel by the top BN layer (Figure 1e). Afterwards, the BN/MoSe$_2$ stack is aligned and released onto the metalized BN support to complete the channel encapsulation with the Hall contacts wired outwards (Figure 1f). A second thermal annealing at 350 °C is necessary to remove possible bubbles and strain introduced during transfer. Finally, a local top gate is deposited to accomplish the entire fabrication (Figure 1g). In this specific device, the channel thickness ($t$) is about 3.5 nm (5-layer) as estimated by atomic force microscopy (AFM) and the top BN that serves as gate dielectric is 38 nm. The MoSe$_2$ channel is also verified by Raman characterization (Figure 1h); three characteristic peaks located at 239.2, 284.7 and 351.8 cm$^{-1}$ are seen, which are assigned as the out-of-plane $A_{1g}$, in-plane $E_{2g}^1$, and $B_{2g}^1$ (inactive in bulk) modes, respectively.[23,24]

Figure 2 shows the electrical performance of the encapsulated MoSe$_2$ FET. The device displays unbalanced ambipolar behavior in the transfer curves (Figure 2a), exhibiting strong n-type but weak p-type conduction at all measured $T$s. The on/off ratio reaches about $10^9$ at 300 K and further increases to $10^{10}$ at 8 K, superior to devices supported by SiO$_2$ substrates.[16] Remarkably, the interfaces of devices are ultraclean and feature an extremely low density of residual traps as compared to conventional unencapsulated or SiO$_2$ supported counterparts, because the transfer curves exhibit hardly appreciable scanning hystereses at all $T$s from 300 and 8 K, as the local top gate is scanned forward and backward between -10 and 12 V. These scanning hystereses ($\Delta V_H$) become discernible only in the enlarged plots, which are estimated to be 12 and 4.5 mV at 300 K and 8 K, respectively (insets of Figure 2a). It is widely recognized that the scanning hystereses originate mostly from the repeated processes of capture and release of the interfacial traps located at the dielectrics and substrates.[25] Since the both interfaces of our devices are contacted with the surface-saturated BN dielectrics, the density of overall interfacial traps can then be minimized. By assuming full rates in the capture and release of traps during gate scan and considering the realistic value of coupling capacitance $C_{tg} = 9.3 \times 10^{-8}$ F·cm$^{-2}$, the densities of traps ($N_t$s) are estimated to be 7.0 and 2.6×10$^9$ cm$^{-2}$ at 300 and 8 K, respectively, using the relation $N_t \sim (C_{tg} \times \Delta V_H)/q$. The reduced level of active traps at low $T$ can be attributed to the suppression of traps on deep energy levels due to lowered thermal excitation. These values are superior to that in the passivated Si/SiO$_2$ interfaces (typically ~10$^{10}$ cm$^{-2}$). The overall low trap levels verify the ultrahigh interface quality of the bottom BN support after thermal annealing, which eliminates the possible risk of contamination during lithography.

The high interface quality is also corroborated by the steep subthreshold swings (SS), which approach the Boltzmann limit.[26] Figure 2b shows the SS values extracted at different drain-source current ($I_{ds}$) levels, where a low SS value of 65 mV/dec spans three orders of current in magnitude when $I_{ds}$ is below 10$^{-10}$ A. We note that it is quite nontrivial to observe a near-Boltzmann-limit SS in devices with thick dielectrics because of the presence of generally large trap capacitance ($C_{trap}$) in conventional devices. The switching of current in FETs is associated with thermionic injection of electrons over an energy barrier and thus SS can be expressed as [27,28]

$$\text{SS} = \left(1 + \frac{C_{trap}+C_d}{C_{ox}}\right) \ln 10 \; \frac{kT}{q} \cong \left(1 + \frac{C_{trap}}{C_{ox}}\right) 60 \; \text{mV/dec} \tag{1}$$

where $kT/q$ is the thermal voltage and $C_d$ is the depletion capacitance that is negligible for 2D channels. Hence, SS in 2D FETs is mainly determined by the ratio between $C_{trap}$ and $C_{ox}$. Only in devices with ultrathin high-k gate dielectric (i.e., large $C_{ox}$)[29] or extremely clean interfaces ($C_{trap} \sim 0$) can the near-Boltzmann-limit SS values be seen. By assuming $C_{trap} = D_{it}q$, where $D_{it}$ denotes the interface trap density, one can roughly estimate $D_{it}$ with the formula[30–33]

$$D_{it} = \frac{C_{tg}}{q}\left(\frac{q \cdot \text{SS}}{kT \ln 10} - 1\right). \tag{2}$$

Using the SS value, we extract a low effective $D_{it}$ of 4.8×10$^{10}$ cm$^{-2}$ eV$^{-1}$, far below those reported at the interfaces of SiO$_2$[15,25] and high-k dielectrics.[31] Together with the net $N_t$ value extracted from the scanning hystereses, we estimate a narrow bandwidth of 290 meV for trap levels. Surface morphologies on the channel areas from AFM (Figure 2c) also confirm the ultraclean interface of the bottom BN support where no appreciable traces of resist residues are found. All the characterization techniques support the high interface quality.

Also, the FET exhibits excellent Ohmic contacts, where a linear relation between $I_{ds}$ and drain-source bias ($V_{ds}$), together with high $I_{ds}$ magnitudes, is seen in the output curves at low $V_{ds}$ regime (Figure 2d). Such a linear $I_{ds}$-$V_{ds}$ relation persists even down to 8 K (inset of Figure 2d), suggesting the extremely high quality of the van der Waals contacts between Au and MoSe$_2$.

The simultaneously high interface and contact qualities allow us to assess the intrinsic electronic transport of the few-layer MoSe$_2$. By employing the 4-probe method,[34] we first extracted the channel conductivity, $\sigma = (I_{ds}/\Delta V)(L_{in}/W)$, as a function of $V_{tg}$ and $T$, where $\Delta V$ is the voltage drops in the inner contact pairs, $L_{in}$ = 3 μm is the distance of the inner probes and $W$ = 8 μm is the width of the channel region (Figure 1f). Figure 3a shows $\sigma$ versus $V_{tg}$ at various $T$s ranging from 300 to 150 K. The device displays a typical insulating state at low $V_{tg}$ regime and becomes metallic at high $V_{tg}$ regime, as a natural consequence of modulation of carrier density upon electric gating field. The $\sigma$-$V_{tg}$ slope exhibits a negative correlation with $T$, implying the enhancement of carrier mobility at low $T$. To further check the transition between the two conduction states, we also summarize the $\sigma$-$T$ curves at various $V_{tg}$ values (inset of Figure 3a). Clearly, the $\sigma$-$T$ trend transitions from positive to negative as $V_{tg}$ increases. An insulator-metal transition occurs at $V_{tg} \sim 4.2$ V. Such a gating-induced electronic transition agrees with the observations in other high-quality TMDCs.[35,36]

The dependence of field-effect $\mu$ on $T$ is also carefully analyzed for the few-layer MoSe$_2$ channels, as shown in Figure 3b. For the 3.5 nm (5-layer) sample, $\mu$ reaches 68 and 690 cm$^2$V$^{-1}$s$^{-1}$ at 300 and 8 K, respectively. The $\mu$ values slightly decreases when $t$ is reduced to 3.0 nm (4-layer), where $\mu$ is lowered to 53 and 515 cm$^2$V$^{-1}$s$^{-1}$ at 300 and 8 K, respectively. The high sensitivity of $\mu$ to $t$ is a common character for few-layer TMDCs, which has been attributed to various interface relevant carrier scattering mechanisms,[11] including charged impurities,[37] remote interface phonons, and lattice vacancies.[38] In this case, since the BN encapsulated channel interfaces are ultraclean and exhibit minimized trap states, we deduce that remote interface phonons and lattice vacancies are the primary scattering mechanisms dominating the low-$T$ electronic transport.

Apart from the above two types of primary scattering mechanisms, lattice phonons represent another type, but is activated mainly at high $T$ regimes. A power law $\mu \propto T^{-\gamma}$ is often employed to analyze the trend of $\mu$ around room $T$ to clarify the weights of extrinsic scattering mechanisms,[20,38] because the increase of its weight would lower the value of the exponent $\gamma$ relative to the intrinsically theoretical value.[39] From the $\mu$-$T$ curves of the two MoSe$_2$ samples, we extract a low $\gamma$ of 1.4, as compared to that from the bulk (~2.4). The reduced $\gamma$ value in the few-layer samples further confirms the important roles of remote interface phonons and lattice vacancies that limit the electronic performance. In spite of these limiting factors, the observed magnitudes of $\mu$ at room $T$ are among the best results reported so far,[15,16,40–42] as shown in Figure 3c. The individual $\mu$ values are above the fitted $\mu$-$t$ trend (dashed line in Figure 3c). The outstanding electronic performance is attributed to the ultrahigh interface and contact qualities that benefit from the polymer-free pickup-transfer technique and the van der Waals contacts.

Next, we turn to the electronic characteristics of the van der Waals Au/MoSe$_2$ contacts by analyzing the interfacial Schottky barrier ($\Phi_B$). According to the theory of thermionic emission, $\Phi_B$ can be extracted from the subthreshold current through the 2D thermionic emission equation [43,44]

$$I_{ds} = A^*_{2D} A T^{\frac{3}{2}} \exp\left[-\frac{q}{kT}\left(\Phi_B - \frac{V_{ds}}{n}\right)\right] \quad (3)$$

where $A^*_{2D}$ is the 2D Richardson constant, $A$ is the junction area, $n$ is the ideality factor accounting for barrier lowering by image charges. Because the application of $V_{ds}$ or $V_{tg}$ can result in the variation of the effective barrier height ($\Phi_E$), it is necessary to extract the intrinsic $\Phi_B$ by extrapolating $\Phi_E$ to zero $V_{ds}$ under the flat-band condition. By setting $\Phi_E = (\Phi_B - V_{ds}/n)$, at a fixed $V_{ds}$ one can extract the values of $\Phi_E$ at different $V_{tg}$ from the slopes in the Arrhenius plots of $\ln(I_{ds}/T^{3/2})$ versus $1000/T$ (Figure 4a). For each $V_{ds}$, the corresponding $\Phi_E$ under the flat-band condition ($\Phi_{E,fb}$) can be then obtained through checking the point that deviates the linear trend in the relevant $\Phi_E$ -$V_{tg}$ curve (Figure 4b). For instance, at $V_{ds}$ = 0.1 V the $\Phi_{E,fb}$ and flat-band voltage are estimated to be 50 meV and 4 V, respectively. Figure 4c displays $\Phi_{E,fb}$ at different $V_{ds}$, from which a small value of 68 meV is extracted for $\Phi_B$. This small value also agrees with the Ohmic behavior observed in the van der Waals contacts as illustrated in Figure 2d.

To further shed light into the physics of the van der Waals contacts, we analyze the dependence of three crucial contact quantities (i.e., $R_C$, $\rho_C$, and $R_S$) on carrier density ($n_{2D}$) and $T$. Since the threshold voltage ($V_T$) of FETs is prone to shift with $T$, it is more reliable to estimate $n_{2D}$ by using $n_{2D} = C_{tg} \cdot (V_{tg} - V_T)$ to rule out the effect of $V_T$ shift. The three contact quantities closely correlate with one another following the injection equation $R_C = \sqrt{\rho_C R_S} \coth(L_C/\sqrt{\rho_C/R_S})$,[45] where $L_C$ is the physical length of contacts. Among the three, $R_C$ and $R_S$ can be directly extracted from experiment while $\rho_C$ is derived from the injection equation. Their dependence on $n_{2D}$ is plotted in Figures 4d–4f, respectively. At all measured $T$s from 300 to 8 K, the three contact quantities share a similar negative correlation with $n_{2D}$, that is they decrease as $n_{2D}$ increases, originating from the mechanism of thermally assisted field emission for electron injection,[46] where the tunneling of carriers is facilitated at high $n_{2D}$ levels due to the narrowing of injection barriers.

In contrast to $n_{2D}$, the dependence of the three contact quantities on $T$ is more complicated. From Figures 4d and 4e, one can find that, in contrast to $R_s$, both $R_C$ and $\rho_C$ exhibit less strong dependence on $T$. In particular, they become nearly independent with $T$ below 70 K (Insets of Figures 4d and 4e). This feature is in line with the injection mechanism of thermally assisted field emission, in which the effect of $T$ is much reduced, as compared with the purely thermionic injection mechanism where $\rho_C \propto T^{-3/2} \exp(q\Phi_B/k_B T)$. In general, $\rho_C$ is slightly negatively correlated with $T$ (inset of Figure 4e) because of the weak $T$ dependence in the thermally assisted field emission process, whereas $R_S$ exhibits strongly positive correlation with $T$ (inset of Figure 4f) due to the enhancement of $\mu$ (Figure 3b), arising from the suppression of thermally relevant carrier scattering mechanisms at low $T$. As a result, a nonmonotonic relationship between $R_C$ and $T$ is observed (inset of Figure 4d) because it is determined by the competition between the two opposite trends of $\rho_C$ and $R_S$. As presented in the inset of Figure 4d, at $n_{2D} = 5 \times 10^{12}$ cm$^{-2}$, $R_C$ exhibits a reverse of $T$ correlation at 240 K.

In our devices with a large $L_C$ over 3 μm, the limiting factor in the injection equation $\coth(L_C/\sqrt{\rho_C/R_S}) \sim 1$, which results in a relative concise relation between the three contact quantities: $R_C \sim \sqrt{\rho_C R_S}$. In this case, the trend of $R_C$ with different device parameters (e.g., $n_{2D}$ and $T$) can thus be readily understood in terms of the trends of $\rho_C$ and $R_S$. At 300 K, the best $R_C$ value is as low as 4.5 kΩ·μm ($n_{2D} = 8.2 \times 10^{12}$ cm$^{-2}$), which is 2−3 orders lower in magnitude than those extracted from conventional contacts,[47–50] and is comparable to the results from contacts with ultralow work functions,[51] indicating the excellent quality of the van der Waals contacts.

We have fabricated high-quality MoSe$_2$ FETs with simultaneous van der Waals encapsulation and contacts with a dry pickup-transfer technique. The devices exhibit extraordinary cleanliness at device interfaces, which results in the observation of the intrinsic device properties, including near-Boltzmann-limit subthreshold swings (~65 mV/dec), negligible scanning hystereses (<15 mV), and high room-$T$ carrier mobilities (53−68 cm$^2$ V$^{-1}$ s$^{-1}$). Moreover, the relevant contact physics in the van der Waals contacts are systematically discussed with clarifying the dependence of contact quantities (i.e., $R_C$, $\rho_C$, and $R_S$) on carrier density and temperature. This work provides insightful information on the device physics in BN encapsulated, van der Waals contacted 2D FETs.

**METHODS**

**Electron beam lithography (EBL) and Metallization.** A bilayer of electron resists comprising methyl methacrylate (MMA, EL9) and poly(methyl methacrylate) (PMMA, A4) was adopted in EBL to facilitate the subsequent procedures of patterning and lift-off of the contact electrodes on the bottom BN flakes. After electron exposure, the resist patterns were developed, rinsed, and blow dried with N$_2$. Then, the contact electrodes (3 nm Ni/12 nm Au) were defined with a combination of thermal evaporation deposition and standard lift-off. Finally, the bottom BN flakes with defined electrodes were thermally annealed at 250 °C for 2 h in Ar/H$_2$ atmosphere to remove the possible resist residues.

**Van der Waals assembly.** To fabricate the BN-encapsulated FETs, individual flakes of BN and MoSe$_2$ were initially mechanically exfoliated onto SiO$_2$/Si substrates. A relatively thick BN flake was selected as a top encapsulator and picked up with a polydimethylsiloxane (PDMS) slab coated with poly(propylene carbonate) (PPC) polymer at 40 °C in a nitrogen glovebox. Afterward, the BN/PPC/PDMS trilayer stack was used to pick up $MoSe_2$ flakes and transfer them to the bottom BN encapsulator with predefined contact electrodes to form a more complex hexa-layer stack. Then, the hexa-layer stack was heated at a softening temperature of 120 °C for the PPC layer above to release the encapsulated BN/$MoSe_2$/contacts/BN structure from PDMS. Next, the BN/$MoSe_2$/contacts/BN stack was cleaned with acetone and was thermally annealed at 350 °C for 2 h in Ar/$H_2$ atmosphere to remove the bubbles and strain introduced during transfer. Finally, a metallic top gate and wiring pads (5 nm Ni/50 nm Au) were defined together by standard EBL and metallization.

## Acknowledgements

This work was supported by the National Key R&D Program of China (Grant Nos. 2022YFA1203802 and 2021YFA1202903), the National Natural Science Foundation of China (Grant Nos. 92264202, 61974060, and 61674080), the Innovation and Entrepreneurship Program of Jiangsu Province.

Published at Chinese Physics Letters, 40, 068501 (2023)     DOI: 10.1088/0256-307X/40/6/068501and Banerjee S K 2015 *ACS Nano* **9** 10402

[36] Radisavljevic B and Kis A 2013 *Nat. Mater.* **12** 815

[37] Li S L, Wakabayashi K, Xu Y, Nakaharai S, Komatsu K, Li W W, Lin Y F, Aparecido-Ferreira A and Tsukagoshi K 2013 *Nano Lett.* **13** 3546

[38] Cui X, Lee G H, Kim Y D, Arefe G, Huang P Y, Lee C H, Chenet D A, Zhang X, Wang L, Ye F, Pizzocchero F, Jessen B S, Watanabe K, Taniguchi T, Muller D A, Low T, Kim P and Hone J 2015 *Nat. Nanotechnol.* **10** 534

[39] Fivaz R and Mooser E 1967 *Phys. Rev.* **163** 743

[40] Ghiasi T S, Quereda J and van Wees B J 2018 *2D Materials* **6** 015002

[41] Lee H, Kim J H and Lee C J 2016 *Appl. Phys. Lett.* **109** 222105

[42] Liao W, Wei W, Tong Y, Chim W K and Zhu C 2017 *Appl. Phys. Lett.* **111** 082105

[43] Kim C, Moon I, Lee D, Choi M S, Ahmed F, Nam S, Cho Y, Shin H J, Park S and Yoo W J 2017 *ACS Nano* **11** 1588

[44] Yu L, Lee Y H, Ling X, Santos E J G, Shin Y C, Lin Y, Dubey M, Kaxiras E, Kong J, Wang H and Palacios T 2014 *Nano Lett.* **14** 3055

[45] Murrmann H and Widmann D 1969 *IEEE Transactions on Electron Devices* **16** 1022

[46] Li S L, Komatsu K, Nakaharai S, Lin Y F, Yamamoto M, Duan X and Tsukagoshi K 2014 *ACS Nano* **8** 12836

[47] Yang L, Majumdar K, Liu H, Du Y, Wu H, Hatzistergos M, Hung P Y, Tieckelmann R, Tsai W, Hobbs C and Ye P D 2014 *Nano Lett.* **14** 6275

[48] Ovchinnikov D, Allain A, Huang Y S, Dumcenco D and Kis A 2014 *ACS Nano* **8** 8174

[49] Kumar J, Kuroda M A, Bellus M Z, Han S J and Chiu H Y 2015 *Appl. Phys. Lett.* **106** 123508

[50] Somvanshi D, Ber E, Bailey C S, Pop E and Yalon E 2020 *ACS Appl. Mater. Interfaces* **12** 36355

[51] Ju S, Qiu L, Zhou J, Liang B, Wang W, Li T, Chen J, Wang X, Shi Y and Li S 2022 *Appl. Phys. Lett.* **120** 25350510/14

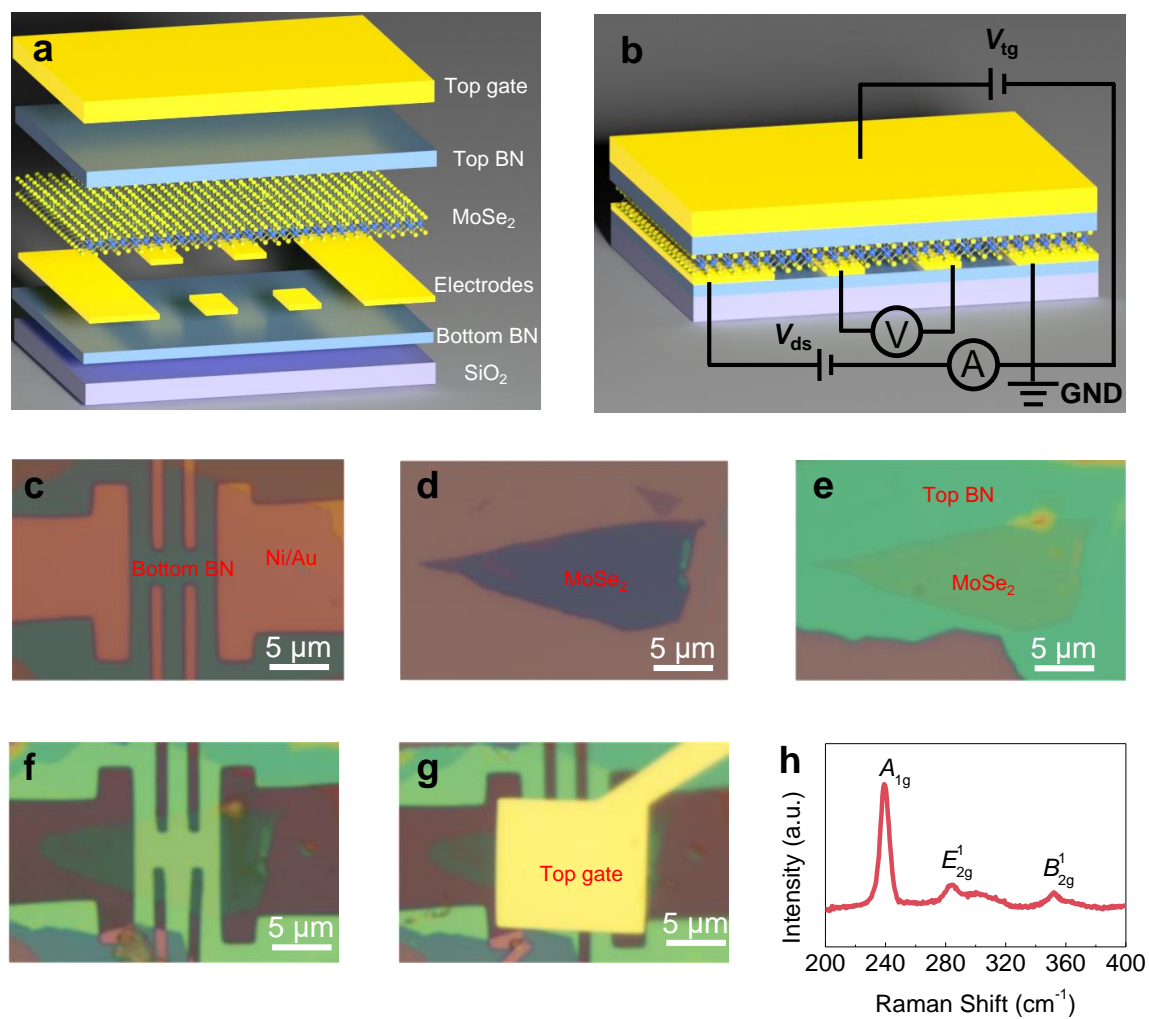

**Figure 1. a**, Schematic illustration for all layers that comprise the BN encapsulated $MoSe_2$ FET. **b**, Final device and wiring configurations. **c-g**, Optical images for (**c**) pre-defined Ni/Au (3/12 nm) electrodes on bottom BN, (**d**) $MoSe_2$ flake exfoliated onto a $SiO_2$/Si substrate, (**e**) $MoSe_2$ flake picked up by top BN, (**f**) fully BN encapsulated $MoSe_2$, and (**g**) as-fabricated device. **h**, Raman spectra for the $MoSe_2$ channel.

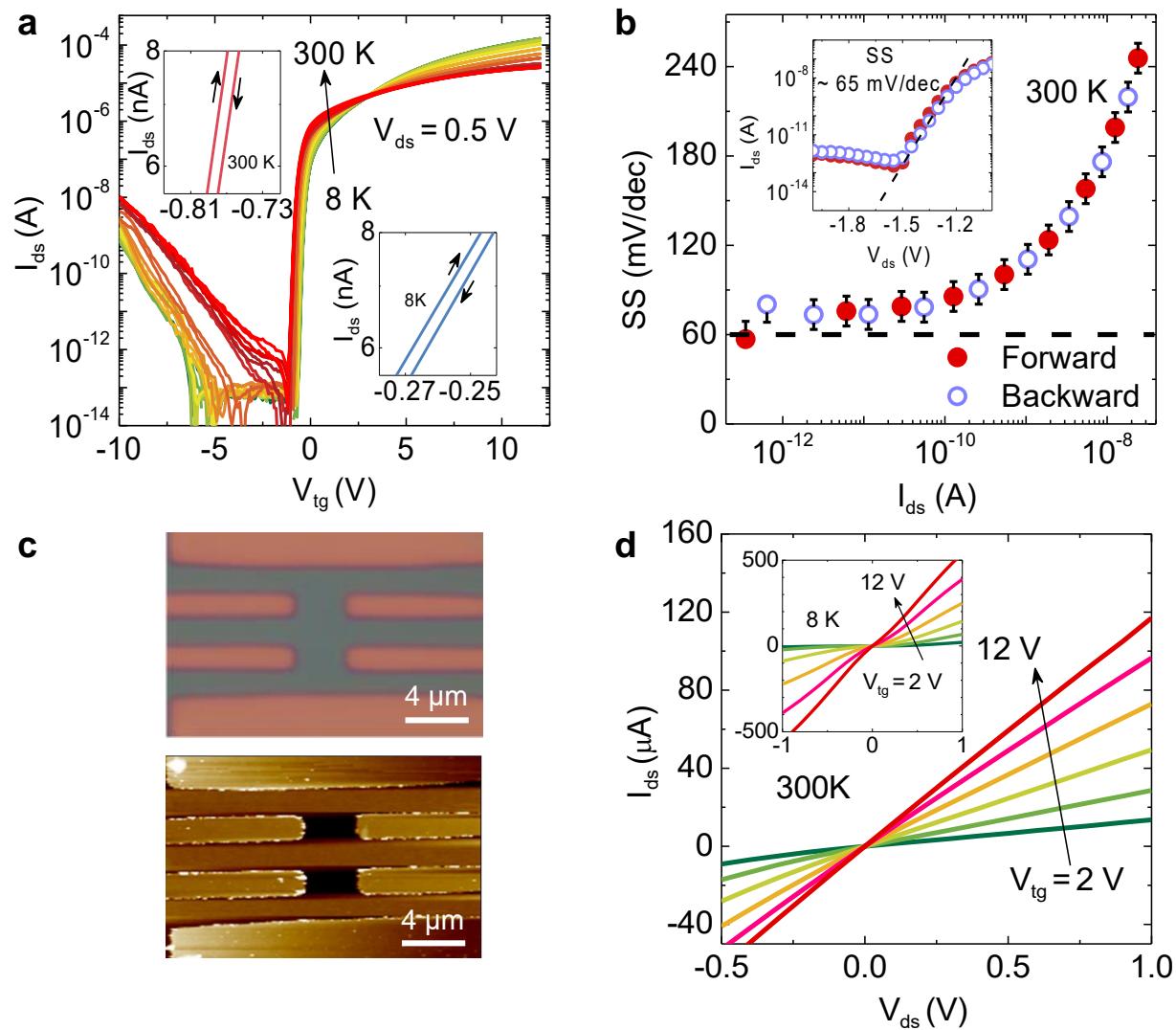

**Figure 2. a**, Transfer characteristics of the MoSe$_2$ FET at different $T$s. Insets: Enlarged plots to show scanning hystereses. **b**, SS versus $I_{ds}$. Inset: $I_{ds}$ versus $V_{ds}$ at 300 K. **c**, Optical and corresponding AFM images for the surface of a metallized BN support. **d,** Output characteristics of the MoSe$_2$ FET at 300 K and 8 K (inset).

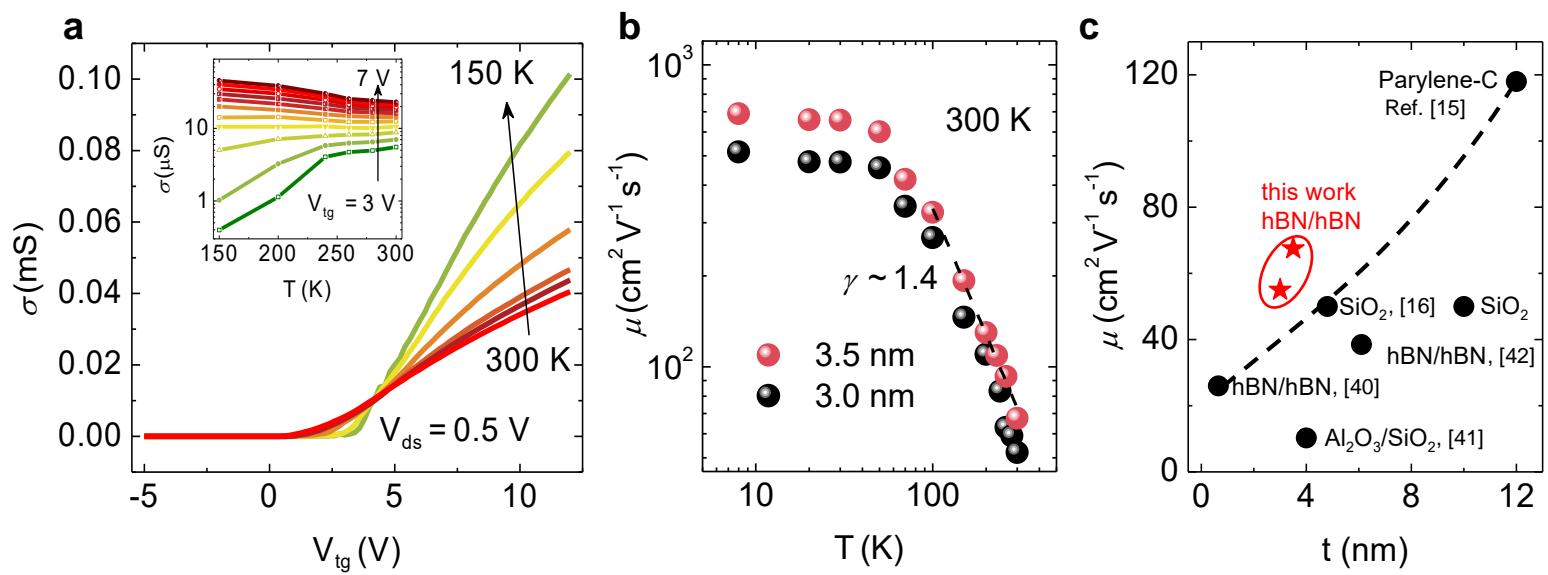

**Figure 3. a**, 4-probe $\sigma$ versus $V_{tg}$ at different $T$s. Insert: $\sigma$ versus $T$ at different $V_{tg}$s. **b**, Dependence of $\mu$ on $T$ for two typical MoSe$_2$ FETs. **c**, Comparison of room-$T$ mobilities of MoSe$_2$ FETs with different dielectrics.

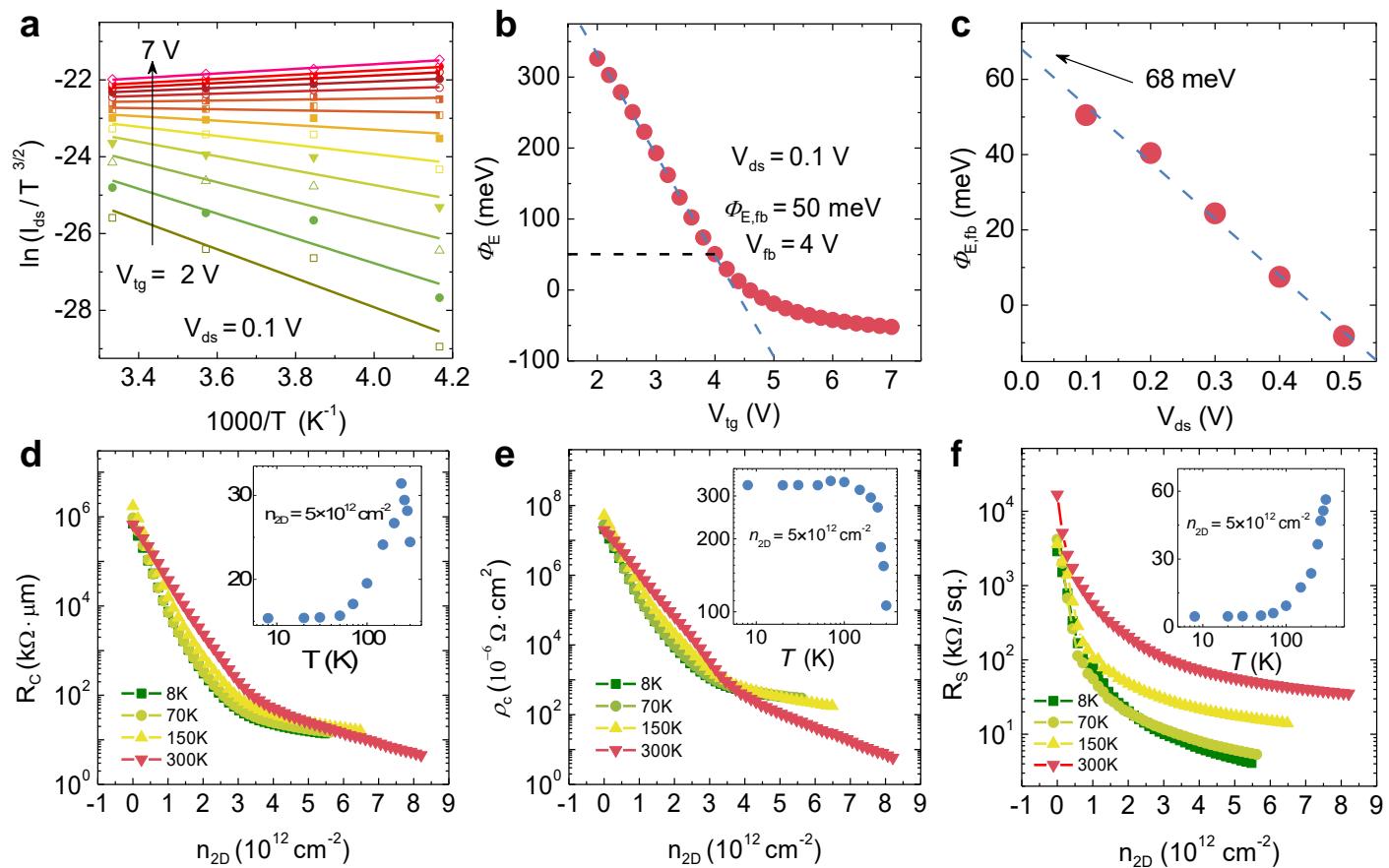

**Figure 4. a**, Arrhenius plots of $I_{ds}/T^{3/2}$ versus $1000/T$ at different $V_{tg}$s. **b**, Extracted $\Phi_E$ values at different $V_{tg}$s. $\Phi_{E,fb}$ is extracted under the flat-band condition. **c**, $\Phi_{E,fb}$ versus $V_{ds}$. **d**, $R_C$ versus $n_{2D}$ (inset: $R_C$ versus $T$). **e**, $\rho_C$ versus $n_{2D}$ (inset: $\rho_C$ versus $T$). **f**, $R_S$ versus $n_{2D}$ (inset: $R_S$ versus $T$).